\documentclass[notitlepage,nofootinbib,preprintnumbers,amssymb,superscriptaddress]{revtex4-1}
\usepackage{amsfonts,amssymb,amsmath,graphicx,color,bm}
\definecolor{ultramarine}{rgb}{0.07, 0.04, 0.56}
\definecolor{cadmiumgreen}{rgb}{0.0, 0.42, 0.24}
\definecolor{indigo(dye)}{rgb}{0.0, 0.25, 0.42}
\usepackage[normalem]{ulem}
\usepackage[linktocpage=true]{hyperref}
\hypersetup{
colorlinks=true,
citecolor=green,
linkcolor=cadmiumgreen,
urlcolor=indigo(dye),
}

\def\[{\begin{equation}}
\def\]{\end{equation}}

\newcommand*{\cH}{{\mathcal{H}}}
\newcommand*{\cF}{{\mathcal{F}}}
\newcommand*{\cG}{{\mathcal{G}}}

\newcommand*{\cJ}{{\mathcal{J}}}
\newcommand*{\cZ}{{\mathcal{Z}}}

\allowdisplaybreaks

\begin{document}

\title{Linear stability of a time-dependent, spherically symmetric background in beyond Horndeski theory and the speed of gravity waves}

\author{S. Mironov}
\email{sa.mironov\_1@physics.msu.ru}
\affiliation{Institute for Nuclear Research of the Russian Academy of Sciences,
60th October Anniversary Prospect, 7a, 117312 Moscow, Russia}
\affiliation{Institute for Theoretical and Mathematical Physics,
MSU, 119991 Moscow, Russia}
\affiliation{NRC, "Kurchatov Institute", 123182, Moscow, Russia}

\author{M. Sharov}
\email{sharov.mr22@physics.msu.ru}
\affiliation{Institute for Nuclear Research of the Russian Academy of Sciences,
60th October Anniversary Prospect, 7a, 117312 Moscow, Russia}
\affiliation{Department of Particle Physics and Cosmology, Physics Faculty, M.V. Lomonosov Moscow State University,
Vorobjevy Gory, 119991 Moscow, Russia}

\author{V. Volkova}
\email{volkova.viktoriya@physics.msu.ru}
\affiliation{Institute for Nuclear Research of the Russian Academy of Sciences,
60th October Anniversary Prospect, 7a, 117312 Moscow, Russia}
\affiliation{Department of Particle Physics and Cosmology, Physics Faculty, M.V. Lomonosov Moscow State University,
Vorobjevy Gory, 119991 Moscow, Russia}

\begin{abstract}
We address a dynamical, spherically symmetric background in beyond Horndeski theory and formulate a set of linear stability conditions for high energy 
perturbation modes in the parity odd sector above an arbitrary solution. In this general setting we derive speeds of propagation in both radial and angular directions for the only dynamical degree of freedom in the parity odd sector. We also briefly comment on the propagation speeds of the parity even modes over a dynamical, spherically symmetric background. In particular, we demonstrate that the class of beyond Horndeski theories, which satisfy the equality of gravity waves' speed to the speed of light over a cosmological background, feature gravity waves propagating at luminal speeds above a time-dependent inhomogeneous background as well.
\end{abstract}

\maketitle

\section{Introduction}\label{sec:intro}

Scalar-tensor theories like Horndeski theories~\cite{Horndeski:1974wa,Deffayet:2011gz,Kobayashi:2011nu} and their generalizations (GLPV or beyond Horndeski~\cite{Zumalacarregui:2013pma,Gleyzes:2014dya,Gleyzes:2014qga}  and DHOST theories~\cite{Langlois:2015cwa,Crisostomi:2016czh,BenAchour:2016fzp}) provide a particularly useful framework for considering theories of modified gravity on general grounds.
The application range of scalar-tensor theories in cosmology is truly impressive: it allows modelling the earliest stages of the Universe, e.g. inflationary epoch or a more exotic Genesis stage,
as well as it is applicable for the most up to date issues related to dark sector, in particular, the late-time accelerated expansion of the Universe (see e.g. Refs.~\cite{Heisenberg:2018vsk,Kobayashi:2019hrl} for reviews).

However, the detection of gravitational wave (GW) signal from a neutron star merger GW170817 and its optical counterpart~\cite{LIGOScientific:2017zic} has severely constrained scalar-tensor theories provided the latter are accountable for modelling Dark Energy~\cite{Ezquiaga:2017ekz,Creminelli:2017sry,Baker:2017hug,Langlois:2017dyl,Kase:2018aps} (see, however, Ref.~\cite{deRham:2018red}). Indeed, the GW speed ($c_{GW}$) is generally modified within scalar-tensor theories as compared to the speed of light ($c_{\gamma}=1$), so that the results of GW170817, i.e.
$|\frac{c_{GW}}{c_{\gamma}}-1| \leq 5 \times 10^{-16}$, rule out a significant number of subclasses of Horndeski theories and their generalisations as Dark Energy candidates or late-time modifications of gravity. In this paper we focus on a subclass of beyond Horndeski theories, whose Lagrangian reads:
\[
\label{eq:lagrangian_intro}
	\begin{aligned}
\mathcal{L} &= F(\pi,X) - K(\pi,X)\Box\pi + G_4(\pi,X)R + G_{4X}(\pi,X)\left[\left(\Box\pi\right)^2-\pi_{;\mu\nu}\pi^{;\mu\nu}\right] \\
&+ G_5(\pi,X)G^{\mu\nu}\pi_{;\mu\nu}-\frac{1}{6}G_{5X}\left[\left(\Box\pi\right)^3-3\Box\pi\pi_{;\mu\nu}\pi^{;\mu\nu}+2\pi_{;\mu\nu}\pi^{;\mu\rho}\pi_{;\rho}^{\;\;\nu}\right] \\
	&- 2 F_4(\pi,X) \left( X \left[\left(\Box\pi\right)^2-\pi_{;\mu\nu}\pi^{;\mu\nu}\right] + \left[\pi^{,\mu} \pi_{;\mu\nu} \pi^{,\nu}\Box\pi -  \pi^{,\mu} \pi_{;\mu\lambda} \pi^{;\nu\lambda}\pi_{,\nu} \right] \right),
	\end{aligned}
\]
where
$F$, $K$, $G_4$, $G_5$ and $F_4$ are arbitrary functions of a
scalar field $\pi$ and $X= -\frac12\: g^{\mu\nu}\pi_{,\mu}\pi_{,\nu}$, and
$\pi_{,\mu}=\partial_\mu\pi$,
$\pi_{;\mu\nu}=\triangledown_\nu\triangledown_\mu\pi$,
$\Box\pi = g^{\mu\nu}\triangledown_\nu\triangledown_\mu\pi$,
$G_{iX}=\partial G_i/\partial X$.
The GW speed depends on $G_4$, $G_5$ and $F_4$ and the background dynamics in a non-trivial way (see e.g. Refs.~\cite{Heisenberg:2018vsk,Kobayashi:2019hrl}). 
The requirement $c_{GW} = c_{\gamma}$ results in the following relations for Lagrangian functions in eq.~\eqref{eq:lagrangian_intro}~\cite{Ezquiaga:2017ekz,Creminelli:2017sry}
\footnote{Note that there are different sign conventions for function $F_4$ in the Lagrangian, as well as different definitions of the kinetic term $X$. We provide the relations in accordance with the notations in eq.~\eqref{eq:lagrangian_intro}.}
\[
\label{eq:GWcosntraint}
G_5 = 0, \quad F_4 = \frac{G_{4X}}{2 X},
\]
provided that no fine-tuning of the background is involved. 
Let us note in passing that the constraints above are applicable only to the original formulation of (beyond) Horndeski theories, where the photon is minimally coupled to gravity, hence, its speed $c_{\gamma}=1$ (see Refs.~\cite{Mironov:2024idn,Babichev:2024kfo,Mironov:2024umy} for further discussion).

{It was {discussed and} argued already
in e.g. Refs.~\cite{Ezquiaga:2017ekz,Babichev:2017lmw,Bettoni:2016mij} that the relation for speeds $c_{GW} = c_{\gamma}$, guaranteed by constraints~\eqref{eq:GWcosntraint} in the case of Lagrangian~\eqref{eq:lagrangian_intro}, remains valid over general backgrounds, including the case of the inhomogeneous setting.}
{In this paper we 
{explicitly demonstrate that the equality of the speeds indeed holds over a time-dependent, spherically symmetric background.}
To do so we consider a spherically symmetric background of general form with time-dependent metric functions:
\[
\label{eq:backgr_metric}
ds^2 = - A(t,r)\:dt^2 + \frac{dr^2}{B(t,r)}
+ J^2(t,r) \gamma_{ab} dx^a dx^b,
\]
where $\gamma_{ab}=\mbox{diag}(1, \sin^2\theta)$ and $a,b=\theta,\varphi$, and
the background scalar field $\pi = \pi(t,r)$ also depends on time and radial coordinate in an arbitrary way. 
{In order to check} that the GW speed is close or equal to luminal we focus on axial (or parity odd) perturbations and derive the corresponding radial and angular propagation speeds {of the only existing odd-parity degree of freedom (DOF), which can be associated with one of the polarization modes of the graviton. We also formulate}
a set of constraints for the Lagrangian functions, which are sufficient to claim stability of the background solution at least w.r.t. high energy modes {in the parity odd sector}. The outlined stability analysis for linear perturbations follows the steps of the existing studies carried out in the case when the scalar field is linearly time-dependent (i.e. $\pi = q\cdot t + \psi(r)$, where $\psi(r)$ is an arbitrary function) while the background metric remains static~\cite{Ogawa,Takahashi:2016dnv,Babichev:2018uiw,Takahashi,deRham:2019gha,Tomikawa:2021pca,Langlois:2021aji,Takahashi:2021bml,Langlois:2022eta,Nakashi:2022wdg,Noui:2023ksf}. We pay special attention to possible appearance of superluminality as a potentially problematic feature for a homogeneous setting in scalar-tensor theories discussed in different contexts~\cite{Dubovsky:2005xd,Adams:2006sv,Babichev:2007dw},
including non-singular cosmological models in scalar-tensor theories~\cite{Creminelli:2012my,Dobre:2017pnt,Mironov:2019mye,Mironov:2023wxn} and problems with matter coupling~\cite{Easson:2013bda,Mironov:2020mfo,Mironov:2020pqh}.
{As for the polar (or parity even) perturbation modes the situation is more complicated since two even-parity DOFs get considerably mixed, which makes it hard to identify the second polarization mode of the graviton.
Nevertheless, we briefly comment on this issue and provide a preliminary analysis of the even-parity sector, which shows that at least the radial speed of one of the even-parity DOFs coincides with that of the odd-parity DOF. }

In result we formulate a set of linear stability constraints in the odd-parity sector for high energy perturbations above a dynamical, inhomogeneous background~\eqref{eq:backgr_metric} in beyond Horndeski theory~\eqref{eq:lagrangian_intro}.
These stability criteria are sufficient to ensure the absence of ghost and gradient instabilities {among the odd-parity modes} for a wide range of dynamical background solutions, where non-trivial time-dependence is crucial, e.g. in a gravitational collapse. Another valuable result of this paper is as follows: 
{we have verified that the constraint~\eqref{eq:GWcosntraint} guarantees
the equality of speeds of gravitational and electromagnetic waves in beyond Horndeski theory over an arbitrarily time-dependent, spherically symmetric background.}
{ In particular, this result generalizes the finding 
of~\cite{Babichev:2017lmw}, where the luminality of gravitational waves in theories satisfying~\eqref{eq:GWcosntraint} and in a considerably inhomogeneous background configuration was verified for a specific Schwarzschild-de Sitter solution with a linearly time-dependent scalar profile.}

{The paper is organized as follows.} In Sec.~\ref{sec:odd_parity_bigL} we derive the quadratic action for { the odd-parity perturbations} above the time-dependent, spherically symmetric background in beyond Horndeski theory~\eqref{eq:lagrangian_intro} in general form. In Sec.~\ref{sec:odd_parity_stability_conditions} we formulate the set of stability conditions for high energy modes { in the odd-parity sector} (we consider the dipole mode separately in Sec.~\ref{sec:odd_parity_dipole}). We analyze the speed of gravity waves in the case of a constrained Lagrangian~\eqref{eq:GWcosntraint} in Sec.~\ref{sec:supeluminality}. 
{ In Sec.~\ref{sec:even_parity} we present a preliminary analysis of the even-parity sector 
with the main focus on the values of propagation speeds.} 
We briefly discuss the results and conclude in Sec.~\ref{sec:conclusion}.


\section{Odd-parity perturbations over the time-dependent, spherically symmetric background}
\label{sec:odd_parity_start}

In this section we derive a quadratic action for perturbations
above the dynamical background metric $\bar{g}_{\mu\nu}$ given in eq.~\eqref{eq:backgr_metric}. We make use of the Regge-Wheeler formalism~\cite{ReggeWheeler}: perturbations are classified into odd-parity and even-parity type
based on their behaviour under two-dimensional reflection.
With further expansion of perturbations into series of the spherical harmonics
$Y_{\ell m}(\theta,\phi)$ the modes with
different parity, $\ell$ and $m$ do not mix at the linear level, so
it is legitimate to consider them separately. In the present paper we address 
{in detail} only odd-parity modes, since the even-parity sector becomes highly non-trivial in a fully-fledged time-dependent case and we leave its {proper analysis} for future studies {(see, however, Sec.~\ref{sec:even_parity} for a preliminary discussion)}.

The metric perturbations $h_{\mu\nu}=g_{\mu\nu}-\bar{g}_{\mu\nu}$
in the odd-parity sector can be parametrized
as follows:
\begin{equation}
\label{eq:odd_parametrization}
\begin{aligned}
h_{tt}&=0,\quad h_{tr}=0,\quad h_{rr}=0, \\
h_{ta}&=\sum_{l,m}^{}h_{0,lm}(t,r)E_{ab}\partial^{b}Y_{lm}(\theta,\varphi),\\
h_{ra}&=\sum_{l,m}^{}h_{1,lm}(t,r)E_{ab}\partial^{b}Y_{lm}(\theta,\varphi),\\
h_{ab}&=\frac{1}{2}\sum_{l,m}^{}h_{2,lm}(t,r)[{E_a}^c\nabla_c\nabla_bY_{lm}(\theta,\varphi)
+{E_b}^c\nabla_c\nabla_aY_{lm}(\theta,\varphi)],
\end{aligned}
\end{equation}
where $E_{ab} = \sqrt{\det \gamma}\: \epsilon_{ab}$, with
$\epsilon_{ab}$ being a
totally antisymmetric symbol ($\epsilon_{\theta\varphi} = 1$), and $\nabla_a$ is covariant derivative on a 2-sphere. Note that
the scalar field $\pi(t,r)$ does not acquire odd-parity perturbations.

Due to general covariance not all metric variables in eq.~\eqref{eq:odd_parametrization} are physical and some of them can be removed by choosing a specific gauge. Under infinitesimal coordinate transformations $x^{\mu}\to x^{\mu}+\xi^{\mu}$,
where
\begin{equation}
\xi_t=\xi_r=0, \quad \xi_{a}=\sum_{l,m}^{}\Lambda_{lm}(t,r){E_{a}}^{b}\nabla_bY_{lm},
\end{equation}
with an arbitrary function $\Lambda_{lm}(t,r)$, the metric variables obey the following transformation rules
\begin{equation}
\label{eq:gauge_transformation}
h_{0} \to h_{0} + \dot{\Lambda} - 2\dfrac{\dot{J}}{J}{\Lambda}, \qquad
h_{1} \to h_{1} + {\Lambda}^{\prime}
- 2\dfrac{J'}{J}{\Lambda},\qquad
h_{2} \to h_{2} + 2{\Lambda},
\end{equation}
where dot and prime denote derivatives w.r.t. time and radial coordinate, respectively, and we dropped the subscripts $\ell, m$ and arguments of the functions for brevity. With an appropriate choice of $\Lambda$ one can eliminate one of the three functions. For example, Regge-Wheeler gauge amounts to choosing $\Lambda=-h_2/2$ so that only $h_0$ and $h_1$ are non-vanishing.
Note that the case of a dipole mode ($\ell=1$) requires different gauge choice since $h_{ab}$ vanishes identically. Hence, in what follows we consider the cases with $\ell\geq2$ and $\ell=1$ separately.
Moreover, in our calculations we set $m = 0$ without loss of generality since the the action for $m \neq 0$ modes takes exactly the same form as that with $m = 0$ thanks to spherical symmetry of the background.

Our calculation strategies below are highly analogous to those carried out in (beyond) Horndeski theory for the static, spherically symmetric case in Refs.~\cite{Kobayashi:odd,MRV1} as well as for linearly time-dependent scalar field in Ref.~\cite{Ogawa}. In what follows we compare and contrast our results with these existing ones.

\subsection{Odd-parity perturbations with $\ell \geq 2$ }
\label{sec:odd_parity_bigL}

Substituting the linearized metric~\eqref{eq:backgr_metric} into the action~\eqref{eq:lagrangian_intro} and adopting the Regge-Wheeler gauge choice ($h_2=0$) we arrive to the quadratic action for odd-parity modes:
\[
\label{eq:action_odd}
S^{(2)}_{odd} = \int \mbox{d}t\:\mbox{d}r \left[ A_1 h_0^2+A_2 h_1^2+A_3 \left( {\dot h_1}^2-2 {\dot h_1} h_0'+h_0'^2+\frac{4 J'}{J} {\dot h_1} h_0
+ \frac{4 \dot{J}}{J} h_1 h_0' \right)
+  A_4 h_1 h_0\right],
\]
where we have integrated over angles and taken up the notations of Refs.~\cite{Ogawa,Takahashi}
\begin{subequations}
\label{eq:A1A2A3}
\begin{align}
& A_1 = \frac{\ell (\ell+1)}{J^2}
\left[\frac{d}{d r}\left(
J\: J' \sqrt{\frac{B}{A}}{\cal H}\right)
+\frac{(\ell-1)(\ell+2)}{2\sqrt{AB}}{\cF}
\right], \\
& A_2 =  \frac{\ell (\ell+1)}{J^2} \left[ \frac{d}{d t}\left(
J\: \dot{J} \sqrt{\frac{B}{A}}{\cal H}\right) - {(\ell-1) (\ell+2)}\frac{\sqrt{AB} }{2}{\cG}\right],\\
& A_3 = \dfrac{\ell(\ell+1)}{2} \sqrt{\frac{B}{A}} {\mathcal{H}},\\
& A_4 = \frac{\ell(\ell+1)}{J^2}\sqrt{\frac{B}{A}}
\left[(\ell-1)(\ell+2)  \cJ - 4 \cH J' \dot{J}\right] ,
\end{align}
\end{subequations}
where the background equations were used to simplify the expressions
\footnote{This is the only place where the background equations were utilized. Since in this work we do not consider any specific solution, we do not provide the equations of motion as their explicit form is not particularly illuminating.}.
As compared to the static case the coefficients $A_i$ gained additional contributions due to time-dependence of the metric.
The coefficients $\cF$, $\cG$, $\cH$ and $\cJ$ in eqs.~\eqref{eq:A1A2A3} also get modified as opposed to the static case (cf. Refs.~\cite{Ogawa,MRV1}):
\begin{subequations}
\label{eq:coeff_HGFJ}
\begin{align}
& \cF = 2 \left[G_4 + X G_{5X} \left( \frac{\dot{\pi}}{2 A}\frac{\dot{B}}{B} - \frac{1}{\pi'} \frac{d}{d r}\left(-\frac12 B \pi'^2 \right) \right) - G_{5\pi}\left(X - \frac{\dot{\pi}^2}{ A}\right) + \frac{\dot{\pi}^2}{2 A} (-2 G_{4X} + 4 F_4 X) \right], \\
& \cG = 2 \left[G_4 - 2 G_{4X} \left(X - \frac{\dot{\pi}^2}{2 A}\right)
+G_{5\pi} \left(X - \frac{\dot{\pi}^2}{ A}\right)
 +X G_{5X} \left(\frac{A'}{2 A} B \pi' + \frac{\dot{A} \dot{\pi}}{2 A^2} - \frac{\ddot{\pi}}{A} \right)
+ 4  F_4 X \left(X - \frac{\dot{\pi}^2}{ 2 A}\right)\right], \\
& \cH = 2 \left[G_4 - 2X G_{4X} +X\left(\left(\frac{B J' \pi'}{J} - \frac{\dot{J} \dot{\pi}}{A J} \right) G_{5X} +G_{5\pi}\right) + 4 X^2F_4 \right] , \\
& \cJ = 2 \dot{\pi} \pi' ( G_{4X} -  G_{5\pi} - 2 X F_4 )
- X G_{5X} \left(2 \dot{\pi}' - \frac{A'\dot{\pi}}{A} + \frac{\dot{B}\pi'}{B}\right),
\end{align}
\end{subequations}
where based on eq.~\eqref{eq:backgr_metric}
\[
X = \frac{\dot{\pi}^2}{2 A} - \frac{1}{2}B {\pi'}^2.
\]
Note that in comparison with the case of a static metric the action~\eqref{eq:action_odd} involves a new mixed term $h_1h_0'$, while in other regards the action is similar to its static counterpart. Despite this fact, further analysis is analogous to the case with static metric in Refs.~\cite{Ogawa,Takahashi} as we show below.

Similarly to the static case the action~\eqref{eq:action_odd} does not involve $\dot{h}_0$, hence, $h_0$ is manifestly non-dynamical. However, variation of the action w.r.t. $h_0$ results in the constraint equation, which contains $h_1'$.
So in order
to avoid solving differential equation for $h_0$ let us rewrite eq.~\eqref{eq:action_odd} as
\begin{multline}
	\label{eq:action_odd_square}
S^{(2)}_{odd} = \int \mbox{d}t\:\mbox{d}r \left[
\left( A_1 - \frac{2}{J^2}\frac{d}{dr}\left[ A_3 {J'}{J} \right] \right)h_0^2
+ \left( A_2 - \frac{2}{J^2}\frac{d}{dt}\left[ A_3 {\dot{J}}{J} \right]\right) h_1^2
\right.\\\left.
+ A_3 \left[ {\dot h_1}- h_0' + 2 \left(\frac{J'}{J}h_0 -\frac{\dot{J}}{J}h_1\right)\right]^2
+  \left(A_4 + 8 A_3 \frac{\dot{J}J'}{J^2}\right) h_1 h_0 \right],
\end{multline}
and introduce a Lagrange multiplier $Q$ as follows:
\begin{multline}
	\label{eq:action_odd_Q}
S^{(2)}_{odd} = \int \mbox{d}t\:\mbox{d}r \left[
\left( A_1 - \frac{2}{J^2}\frac{d}{dr}\left[ A_3 {J'}{J} \right] \right) h_0^2
+ \left( A_2 - \frac{2}{J^2}\frac{d}{dt}\left[ A_3 {\dot{J}}{J} \right]\right) h_1^2
\right.\\\left.
+ A_3 \left( 2 Q \left[ {\dot h_1}- h_0' + 2 \left(\frac{J'}{J}h_0 -\frac{\dot{J}}{J}h_1\right)\right] - Q^2 \right)
+  \left(A_4 + 8 A_3 \frac{\dot{J}J'}{J^2}\right) h_1 h_0 \right].
\end{multline}
It is straightforward to verify that the initial action~\eqref{eq:action_odd_square} gets restored as soon as one makes use of the corresponding equation of motion for $Q$ in eq.~\eqref{eq:action_odd_Q}.
At this point integration by parts in eq.~\eqref{eq:action_odd_Q} enables one to shift all derivatives onto $Q$, thus, making both $h_0$ and $h_1$ non-dynamical variables. Then variation of action~\eqref{eq:action_odd_Q} w.r.t. $h_0$ and $h_1$ gives the following constraint equations, respectively:
\begin{subequations}
\label{eq:constraints}
\begin{align}
& 2 \left(A_1  - \frac{2}{J^2}\frac{d}{dr}\left[ A_3 {J'}{J} \right] \right) h_0 + \left(A_4 + 8 A_3 \frac{\dot{J}J'}{J^2}\right) h_1
+ 4 A_3 \frac{J'}{J} Q + 2 \frac{d}{dr}\left[A_3 Q\right] = 0, \\
& 2 \left(A_2 - \frac{2}{J^2}\frac{d}{dt}\left[ A_3 {\dot{J}}{J} \right] \right) h_1 + \left(A_4 + 8 A_3 \frac{\dot{J}J'}{J^2}\right) h_0 - 4 A_3 \frac{\dot{J}}{J} Q - 2 \frac{d}{dt}\left[A_3 Q\right] = 0.
\end{align}
\end{subequations}
Solving eqs.~\eqref{eq:constraints} for $h_0$ and $h_1$ in terms of $Q$ and its derivatives, and substituting the result back into the action~\eqref{eq:action_odd_Q}, we arrive to the quadratic action for the single dynamical DOF -- $Q$:
\begin{multline}
\label{eq:action_odd_final}
S^{(2)}_{odd} = \int \mbox{d}t\:\mbox{d}r\:\sqrt{\frac{B}{A}} J^2
\frac{\ell(\ell+1)}{2(\ell-1)(\ell+2)}\cdot
\left[\frac{1}{A} \frac{\mathcal{F}\mathcal{H}^2}{\mathcal{G} \mathcal{F} + \frac{B}{A} \cdot\cJ^2 } \dot{Q}^2 -
\frac{ B \cdot \mathcal{G}\mathcal{H}^2}{ \mathcal{G} \mathcal{F} + \frac{B}{A} \cdot\cJ^2} (Q')^2 \right.\\ \left.
+ 2 \frac{B}{A} \frac{\cJ\mathcal{H}^2}{ \mathcal{G} \mathcal{F} + \frac{B}{A} \cdot\cJ^2 } Q'\dot{Q}
-\frac{\ell(\ell+1)}{J^2}\cdot \mathcal{H} Q^2 - V Q^2 \right] \; ,
\end{multline}
where the "potential" $V$ 
{reads
\begin{multline}
\label{eq:V}
V = \frac{2 \cH}{J^2} - \cH \frac{d}{dr}\left[J^2 \sqrt{\frac{B}{A}} \frac{\cH^2}{{ \mathcal{G} \mathcal{F} + \frac{B}{A} \cdot\cJ^2}} \left( B\: \cG \cdot\frac{d}{dr}\left[\sqrt{\frac{A}{B}}\frac{1}{J^2\cH}\right] - 
\frac{B}{A} \cJ \cdot \frac{d}{dt}\left[\sqrt{\frac{A}{B}}\frac{1}{J^2\cH}\right]\right)\right] \\
+ \cH \frac{d}{dt}\left[J^2 \sqrt{\frac{B}{A}} \frac{\cH^2}{{ \mathcal{G} \mathcal{F} + \frac{B}{A} \cdot\cJ^2}} \left( \frac{B}{A} \cJ \cdot\frac{d}{dr}\left[\sqrt{\frac{A}{B}}\frac{1}{J^2\cH}\right] + 
\frac{1}{A} \cF \cdot \frac{d}{dt}\left[\sqrt{\frac{A}{B}}\frac{1}{J^2\cH}\right]\right)\right]
.
\end{multline}
Let us note that $V$ governs the behaviour of low energy modes (sometimes referred to as tachyonic instabilities). 
In the following analysis we focus on high energy modes, which are the most harmful ones since they ruin the background solution immediately. In contrast, low energy modes require a finite period of time to build up and, therefore, are less of a problem since the solution might significantly change during the tachyons' development (see e.g. Ref.~\cite{RubakovNEC} for details). The latter means that the explicit form of the background solution is crucial for determining whether the low energy modes are healthy.
This, in turn, indicates that a full-fledged stability analysis is required. 
}

In full analogy with the results for the case of a linearly time-dependent scalar field in Ref.~\cite{Ogawa}, the mixed term $Q'\dot{Q}$ in eq.~\eqref{eq:action_odd_final} identically vanishes upon switching off time-dependence of the background, i.e. $\cJ \to 0$, so that the results for static case gets restored, cf.~\cite{Kobayashi:odd,MRV1}.
In what follows to simplify expressions we introduce a new notation for a denominator present in the derivative terms of the action~\eqref{eq:action_odd_final}:
\[
\label{eq:definition_Z}
\mathcal{Z} =  \mathcal{G} \mathcal{F} + \frac{B}{A}\cdot\cJ^2.
\]

\subsection{Stability conditions for high energy modes}
\label{sec:odd_parity_stability_conditions}

Let us now consider a high momentum regime, which implies that variations of $Q$ in space and time occur at much shorter scales than those characteristic of background values like $\pi(t,r)$ and metric functions. Then {for deriving the dispersion relation} it is legitimate 
{to neglect the terms that are linear in $\dot{Q}$ or $Q'$ in the field equation corresponding to the action~\eqref{eq:action_odd_final}
}. 
Hence, the resulting dispersion relation
reads:
\begin{equation}
\label{eq:dispersion}
\frac{1}{A}\dfrac{\cF\cH^2}{\mathcal{Z}} \omega^2 = B\dfrac{\cG\cH^2}{\mathcal{Z}} {k}^2 + 2 \frac{B}{A} \dfrac{\cJ\cH^2}{\mathcal{Z}} \omega{k}
+ \frac{\ell(\ell+1)}{J^2}\cdot \mathcal{H},
\end{equation}
where we have omitted the "potential" $V(r)$, since its contribution is irrelevant for high energy modes.
To avoid ghost instabilities one has to impose
\[
\label{eq:no_ghost}
\dfrac{\cF}{\mathcal{Z}} > 0,
\]
while ensuring the absence of radial gradient instabilities amounts to requiring
\[
\label{eq:no_gradient}
{\mathcal{Z}} > 0.
\]
The latter guarantees that the frequencies $\omega(k)$ in eq.~\eqref{eq:dispersion} take real values at high momenta.
To ensure stable propagation
in the angular direction, which is governed by the angular part of the Laplace operator with $\ell(\ell+1)$ in eq.~\eqref{eq:dispersion}, one has to require
\[
\label{eq:angular_gradient}
\cH > 0.
\]
Finally, the propagation speeds along the radial and angular direction immediately follow from the dispersion relation~\eqref{eq:dispersion} and read, respectively:
\[
\label{eq:speeds}
c_r^{(\pm)} = \sqrt{\frac{B}{A}}\frac{\cJ}{\cF} \pm \frac{1}{\cF}\sqrt{ \mathcal{Z}}, \qquad
c_{\theta}^2 = \frac{\mathcal{Z}}{\cF\cH},
\]
where $c_r^{+}$ and $c_r^{-}$ correspond to the speed of the outward and inward propagation.
As before one can restore the existing results for stability conditions and propagation speeds for the static background upon taking the limit $\cJ \to 0$ (hence, $\mathcal{Z} \to \cG\cF$), cf. Refs.~\cite{Ogawa,Takahashi,MRV1}.

To sum up, the constraints~\eqref{eq:no_ghost},~\eqref{eq:no_gradient} and~\eqref{eq:angular_gradient} constitute a set of {sufficient }stability conditions for high energy modes.

\subsection{Dipole perturbation: $\ell =1$ }
\label{sec:odd_parity_dipole}

Let us now consider the dipole mode with $\ell=1$. Since the results for modes with $\ell \geq 2$ are in line with those for the linearly-time dependent scalar field and static metric in Refs.~\cite{Ogawa,Takahashi}, we follow the procedure for dipole mode discussed in these works.

It immediately follows from eqs.~\eqref{eq:A1A2A3} that coefficients $A_i$ get simplified in
the case of odd-parity perturbations with $\ell=1$:
\[
\label{eq:simp_A1A2A3}
 A_1 = \frac{2}{J^2} \frac{d}{d r}\left[A_3 J^{\prime} J\right], \quad
 A_2 =\frac{2}{J^2} \frac{d}{d t}\left[A_3 \dot{J} J\right], \quad
  A_4=-8 A_3 \frac{J^{\prime} \dot{J}}{J^2},
\]
and, hence, the quadratic action~\eqref{eq:action_odd_square} takes the form:
\[
\label{eq:action_odd_dipole}
S^{(2)}_{odd} = \int \mbox{d}t\:\mbox{d}r \left[
A_3 \left( {\dot h_1}- h_0' + 2 \left(\frac{J'}{J}h_0 -\frac{\dot{J}}{J}h_1\right)\right)^2  \right].
\]
Now we derive the corresponding equations of motion for $h_0$ and $h_1$, and then by an appropriate choice of $\Lambda(t,r)$ in eq.~\eqref{eq:gauge_transformation} we eliminate $h_1$ (recall that there is still a residual gauge freedom $\Lambda = C(t) J^2$). The resulting equations read:
\begin{align}
\label{eq:dipole_eqs}
\frac{d}{dt}\left[J^2 A_3 \left(-h_0' + 2 \dfrac{J'}{J} h_0 \right)\right] &= 0, \\
(A_3 h_0')' - 2 \dfrac{(A_3 J J')'}{J^2} h_0&= 0.
\end{align}
Upon solving the system~\eqref{eq:dipole_eqs} we obtain its general solution
\begin{equation}
\label{eq:dipole_sol}
h_0= \frac{3 \mathfrak{J}_1}{4 \pi} J^2 \int^r \frac{d r^{\prime}}{J^{4} A_3},
\end{equation}
where $\mathfrak{J}_1$ is an integration constant.
The relation \eqref{eq:dipole_sol} is similar to the results of Refs.~\cite{Ogawa,Takahashi} and has the static limit of Ref.~\cite{Kobayashi:odd}. The time-dependent integration constant that appears after the radial integration in \eqref{eq:dipole_sol} can be eliminated using the residual gauge freedom mentioned above. Then the gauge is completely fixed and $h_0$ is a physical perturbation which corresponds to a metric around a slowly rotating black hole, with $\mathfrak{J}_1$ being associated with its angular momentum. As for the case in general relativity limit, the solution reduces to $h_0=-\mathfrak{J}_1 /\left(4 \pi M_P^2 J\right)$, which matches the Kerr metric expanded to the first order in the angular momentum $\mathfrak{J}_1$.

Thus, in full analogy to the static case and the case with linearly-time dependent scalar field, the dipole mode does not contribute the set of stability constraints~\eqref{eq:no_ghost},~\eqref{eq:no_gradient} and~\eqref{eq:angular_gradient} discussed for model with $\ell \geq 2$.

\subsection{GW170817 constraint for beyond Horndeski theory and luminal propagation of gravity waves}
\label{sec:supeluminality}

Let us now analyse the propagation speeds~\eqref{eq:speeds} of a gravity mode for a specific beyond Horndeski subclass, that is compliant with the constraint~\eqref{eq:GWcosntraint}. 

Since neither of speeds~\eqref{eq:speeds}  depend on either $F(\pi,X)$ or $K(\pi,X)$ functions in the Lagrangian~\eqref{eq:lagrangian_intro}, we do not specify them, and rely only on a specific choice of $G_5(\pi,X)=0$ and $F_4(\pi,X) = G_{4X}(\pi,X)/2X$, while $G_4(\pi,X)$ is still arbitrary. 
Then it immediately follows from eqs.~\eqref{eq:coeff_HGFJ}
and~\eqref{eq:definition_Z} that $\cJ \to 0$, $\cZ \to \cG \cF$ and 
$\cG,\cF, \cH \to 2 G_4$, hence,   
the speeds~\eqref{eq:speeds} take the following explicit form:
\[
\label{eq:speeds_GW}
c^{2}_r = 1,
\quad 
c^2_{\theta} = 1.
\]
Therefore, we have {explicitly} showed that if beyond Horndeski theory is considered as the late-time modification of gravity and complies with constraints~\eqref{eq:GWcosntraint} following from GW170817 observational data, the gravity mode propagates with luminal speed 
{above the arbitrary dynamical, spherically symmetric background.}

{
\section{Even-parity perturbations over the time-dependent, spherically symmetric background: a preliminary analysis}
\label{sec:even_parity}

In this section we present a preliminary analysis of the even-parity perturbations over the background~\eqref{eq:backgr_metric}, focusing on their propagation speeds. 

In a spherically symmetric case the gravitational DOFs corresponding to two polarizations of the graviton get split into parity odd and parity even modes, while the scalar field $\pi$ contributes to the even-parity sector only. 
It has been already shown for (beyond) Horndeski theory with a static scalar field over a static, spherically symmetric background that the two even-parity modes initially corresponding to metric and scalar field perturbations get significantly mixed (see e.g.~\cite{Kobayashi:2014wsa,Mironov:2022rmw}). In particular, the angular speed for neither of even-parity DOFs matches the angular speed of the only odd-parity mode, associated with the graviton~\cite{Mironov:2024pjj}. This fact is a vivid manifestation of the aforementioned mixing of the metric and the scalar field perturbations, which in the end complicates the unambiguous identification of the gravitational DOF in the even-parity sector. 
Let us note, however, that the situation with the radial speeds is more promising: one of the even-parity DOFs propagates with the same speed as the odd-parity DOF at least over the static, spherically symmetric background~\cite{Kobayashi:2014wsa} and, hence, allowing one to identify both gravitational DOFs in a clear way. Therefore, below we aim to find out whether the radial speed of either of even-parity modes matches the one for the odd-parity gravitational DOF $Q$~\eqref{eq:speeds} over a time-dependent, spherically symmetric background~\eqref{eq:backgr_metric}. 

It should be noted that a full-fledged analysis of the even-parity modes is a computationally demanding task even in the static case, let alone in the time-dependent setting~\eqref{eq:backgr_metric} we currently consider. 
Hence, in this section we analyze the propagation speeds of the even-parity modes on a qualitative level which, however, is sufficient to verify the equality of the radial speed for one of the even-parity DOFs and that for odd-parity mode $Q$. The proper study of the even-parity sector will be addressed in future works, including the derivation of corresponding stability conditions which usually introduce non-trivial limitations on the theory as compared to the existing odd sector constraints, see eqs.~\eqref{eq:no_ghost}-\eqref{eq:angular_gradient}.

Generally, the even-parity sector involves 7 modes from the perturbed metric $h_{\mu\nu}$ and 1 mode $\delta\pi$ associated with the perturbed scalar field. Although 3 modes can be removed in result of gauge fixing,
let us consider the quadratic action for the even-parity modes in the following gauge invariant form:
\[
\label{eq:even_action}
S^{(2)}_{even} = \int \mbox{d}t\:\mbox{d}r\: (\varphi^a K_{ab} \varphi^b),
\]
where $\varphi^a$ ($a=\overline{1..8}$) is a vector whose components are given by 8 even-parity modes listed above, while $K_{ab}$ stands for the operator discribing the dynamics of all modes. Upon switching to the Fourier space in~\eqref{eq:even_action} we calculate the rank of
the resulting matrix
$K_{ab}^{Fourier}$, whose entries are given by quadratic polynomials in frequency $\omega$ and space momentum $k$ :
\[
\label{eq:rank1}
\mbox{rank}\left[K_{ab}^{Fourier}\right] = 5.
\] 
This result is natural since as we mentioned above there are 3 out of 8 modes which can be gauged away due to 3 diffeomorphisms.

Let us now check whether among the leftover 5 modes there is a DOF whose radial speed matches that for the odd-parity DOF $Q$, 
see eq.~\eqref{eq:speeds}.
In order to do so we substitute the dispersion relation for $Q$ 
from eq.~\eqref{eq:dispersion} into $K_{ab}^{Fourier}$ and calculate the rank of the resulting matrix in the leading order
in {high} $\omega$ and $k$
\footnote{In particular, in this limit of high $\omega$ and $k$ we neglect the angular terms, which are proportional to $\ell(\ell+1)$ (these terms define the propagation speed in the angular direction).}.
We see that the rank has decreased by 1:
\[
\label{eq:rank1}
\left.\mbox{rank}\left[K_{ab}^{Fourier}\right]\right|_{{dispersion \; relation}} = 4,
\]
which indicates that there is indeed one even-parity DOF which propagates in the radial direction with the exact same speed $c_r^{(\pm)}$ from eq.~\eqref{eq:speeds} as the odd-parity DOF $Q$ does. 
Hence, one may attribute this even-parity DOF to the second polarization mode of the graviton at least in the context of propagation in the radial direction.
{Let us note that with the relation~\eqref{eq:GWcosntraint} this radial speed for the gravitational mode equals to that of light, see eq.~\eqref{eq:speeds_GW}.}

As it was mentioned above the situation with the angular speeds is not really informative for identifying the gravitational modes since similarly to a static case we do not expect the angular speed of either of even-parity modes to coincide with the angular speed of the odd-parity mode. 
The latter fact makes it reasonable to rely on the odd-parity sector when it comes to identifying and analyzing the speed of GW over the inhomogeneous backgrounds like the spherically symmetric one.

}

\section{Discussion and conclusions}
\label{sec:conclusion}

In this paper we have considered the behaviour of the odd-parity perturbations in beyond Horndeski theory~\eqref{eq:lagrangian_intro} above a time-dependent, spherically symmetric background~\eqref{eq:backgr_metric}. The quadratic action for perturbations~\eqref{eq:action_odd_final} was derived in full generality, hence, it has a potentially wide scope of applications across various fields, such as the dynamical scenarios in black hole physics and the cosmological settings, where inhomogeneity is essential. 
{In particular, we have explicitly formulated a set of constraints for Lagrangian functions, which is sufficient to ensure that there are no pathological DOFs among the high energy modes in the odd-parity sector:}
\begin{eqnarray}
\label{stability_G}\nonumber
\mbox{No ghosts:} &\mathcal{F} > 0, \\
\label{stability_F}\nonumber
\mbox{No radial gradient instabilities:} &\mathcal{G} > - \dfrac BA \dfrac{\cJ^2}{\cF}, \\
\label{stability_H}\nonumber
\mbox{No angular gradient instabilities:} &\mathcal{H} > 0.
\end{eqnarray}
{The propagation speeds of the odd-parity gravity mode in both radial and angular direction were also derived and given in eqs.~\eqref{eq:speeds}.}
{Moreover, we have also briefly addressed the even-parity sector. In particular, we showed that one of the even-parity DOFs propagates in the radial direction
with the same speed as the odd-parity DOF.}

Another valuable finding of the paper is related to the case when beyond Horndeski theory plays the role of Dark Energy or is responsible for the modification of gravity at late times and, hence, the theory can be constrained by the multi-messenger speed test GW170817. 
We have explicitly verified that the original 
condition~\eqref{eq:GWcosntraint}
{guarantees that the gravitational and electromagnetic waves propagate at the same speed over
an arbitrary dynamical, spherically symmetric background.}
{The latter conclusion is valid provided one defines the GW propagation speed over the inhomogeneous background
based on the properties of the odd-parity DOF, which can be unequivocally associated with the GW.   }

\section*{Acknowledgements}
The authors are greatful to Eugeny Babichev for valuable discussions.
The work on this project
has been supported by Russian Science Foundation grant № 24-72-10110,
\href{https://rscf.ru/project/24-72-10110/}{  https://rscf.ru/project/24-72-10110/}.


\end{document}